\documentclass[review]{elsarticle}

\usepackage{graphicx}
\usepackage{amsmath}
\usepackage{mathtools}
\usepackage{setspace}
\usepackage{etex}
\usepackage{hyperref}
\usepackage{amsmath}
\usepackage{float}
\usepackage{morefloats}
\usepackage{adjustbox}
\usepackage{rotating}
\usepackage{pdflscape}
\usepackage[toc,page]{appendix}
\usepackage{lineno,hyperref}
\modulolinenumbers[5]

\begin{document}

\begin{frontmatter}

\bibliographystyle{elsarticle-num}

\title{{Index of refraction, Rayleigh scattering length, and Sellmeier coefficients in solid and liquid argon and xenon}}
\author[rvt]{Emily Grace\corref{cor1}}
\ead{Emily.Williams.2012@live.rhul.ac.uk}

\author{Alistair Butcher}

\author{Jocelyn Monroe}

\author{James A. Nikkel}
\address{Royal Holloway, University of London,
Department of Physics,
Egham, Surrey
TW20 0EX,
UK}

\cortext[cor1]{Corresponding author}

\begin{abstract}
Large liquid argon detectors have become widely used in low rate experiments, including dark matter and neutrino research. However, the optical properties of liquid argon are not well understood at the large scales relevant for current and near-future detectors.The index of refraction of liquid argon at the scintillation wavelength has not been measured, and current Rayleigh scattering length calculations disagree with measurements.  Furthermore, the Rayleigh scattering length and index of refraction of solid argon and solid xenon at their scintillation wavelengths have not been previously measured or calculated.  We introduce a new calculation using existing data in liquid and solid argon and xenon to extrapolate the optical properties at the scintillation wavelengths using the Sellmeier dispersion relationship.  
\end{abstract}


\end{frontmatter}

%\linenumbers

%%%%%%%%%%%%%%%%%%%%%%%%%%%%%%%%%%%%%  Introduction
\section{Introduction} %%%%%%NEW!!

%%%%BeginNew
Liquid nobles such as argon and xenon are used many particle detector experiments including neutrino detectors and low-background dark matter detectors. This family of detectors relies on the scintillation light produced by nobles when exposed to external radiation. Understanding such signal in the detectors relies on a precise optical model to simulate the path of the scintillation light between production and detection in the thermal conditions of the detector medium. Key ingredients of this optical model are the index of refraction $n$ and the scattering length, which depends strongly on $n$. The index of refraction has been measured at the scintillation wavelength of 178 $nm$ in liquid xenon, but not yet in liquid argon. These properties are not well known, and indeed calculations differ from measurements by up to 30\%.  The state of current knowledge about the index of refraction and scattering length in argon and xenon is reviewed in section 1.1.. This difference is more impactful in larger detectors.

Knowledge of the scattering length in liquid nobles is becoming increasingly important as detectors get larger.  In the past decade most liquid argon detectors were small prototypes, where the optical path length was often much shorter than the scattering length and therefore scattering was not a significant effect. In constrast, current and planned detectors are large. Presently constructed are the DarkSide 50 (50 $kg$) \cite{Agne:2016}, ArDM (1 ton) \cite{Rubb:2006A}, MicrobooNE (170 tons) \cite{Jone:2013A}, ICARUS (760 tons) \cite{Anto:2012A}, DEAP3600 (3600 kg), and MiniCLEAN(360 kg). Planned are protoDUNE (770 tons) \cite{Char:2016A}, DarkSide 20k( 20k kg) \cite{Davi:2016A}, and DUNE (17,000 tons) \cite{Acci:2016A}. The drift length in these detectors range from 1 meter to 8 meters. 

\subsection{Survey of Previous Literature}%%%%%%%%%%NEW!!

The existing measurements of the index of refraction in argon and xenon are summarized in section 1.1.1.   Historically, measurements of the index of refraction at temperatures at or above the triple point have been used in calculations to predict the value below the triple point, as described in Section 1.1.2.  These calculations predict the wavelength dependence, which has been used in experiment simulations \cite{Ferm:2016A}\cite{Agos:2002A} to model the propagation of photons produced at the scintillation wavelengths in liquid noble targets.

\subsubsection{Measurements}

\begin{itemize}
\item 
Sinnock and Smith~\cite{Sinn:1969A} measured the index of refraction as a function of wavelength at temperatures between 90 $K$ and 20 $K$ in argon and at temperatures between 80 $K$ and 178 $K$ in xenon. These measurements were made in the wavelength range 350-650 $nm$.  The typical experimental error reported is $\pm 0.5\%$. (Data from \cite{Sinn:1969A} is shown in Figures~\ref{fig:ArIorErr1} and ~\ref{fig:XeIorErr1}.)
 
 \item
 Bideu-Mehu \emph{et al.}~\cite{Bide:1981A} measured the index of refraction of room temperature argon and xenon gas between the wavelengths of 140 $nm$ and 174 $nm$ and used these values to find the Sellmeier coefficients for the gas based Sellmeier \cite{Born:1999A} equation. The typical experimental error reported is $\pm 0.1\%$ 
 
\item
Ishida \emph{et al.}~\cite{Ishi:1997A} measured the attenuation length of liquid xenon and argon at the wavelengths of 178 $nm$ and 128 $nm$ respectively. They found values of $66\pm3~cm$ for argon at $87~K$ and $29\pm2~cm$ for xenon at $196~K$.

\item
Solovov \emph{et al.} \cite{Solo:2004A} measured the index of refraction and attenuation length of liquid xenon at the triplet point and obtained a value of 1.69$\pm$0.02 for the index of refraction and 36$\pm2cm$ for the attenuation length.

\item
The ArDM collaboration published an in situ measurement of the attenuation length of liquid argon in the detector. This yield a measurement of 52$\pm$7$cm$ \cite{Calv:2016A}.

\end{itemize}

\subsubsection{Calculation}

Seidel \emph{et al.} calculated the Rayleigh scattering length for liquid argon and xenon. Seidel's calculated values were 90 $cm$ for argon and 30 $cm$ for xenon (the authors did not include the error on their calculation). The calculated Rayleigh scattering length agrees within errors with the measured values for xenon from Ishida et al.  The calculation is robust in the sense that the xenon value was calculated using the measured value of the index of refraction in liquid xenon, with no extrapolation in temperature or pressure. In the case of argon, Seidel \emph{et al.} \cite{Seid:2002A} used STP gas data from from Bideu-Mehu \emph{et al.} \cite{Bide:1981A} to extrapolate the index of refraction at  the scintillation wavelength. This was value was adjusted according to the density change from liquid to solid, but any temperature dependence was neglected, a decision made based on the gas measurements by Achtermann \emph{et al.} \cite{Acht:1993A}. 

\section{Rayleigh Scattering Length Calculation Dependence on Index of Refraction}%%%%%%%%%%%%%%%%NEW!!

In the following calculation, the temperature dependence is allowed.  Similarly to Seidel et al., we then fit the temperature and density-controlled data from Sinnock and Smith [9] to find the Sellmeier coefficients, which enter the calculation of the index of refraction and thereby the Rayleigh scattering length.

%The following is the theoretical frame work used to make the calculations using the temperature and density controlled data from Sinnock and Smith~\cite{Sinn:1969A}.

Rayleigh scattering is the process of light elastically scattering off of particles smaller than the wavelength of light. The length of travel for a photon through a medium before Rayleigh scattering is strongly dependent on the wavelength of the light as well as the optical properties of the material. The Rayleigh scattering equation for liquids and solids is

\begin{equation}
l^{-1} = \frac{16\pi^3}{6\lambda^4}\left[kT\rho^2\kappa_{T}\left(\frac{(n^2 - 1)(n^2 +2)}{3}\right)^2\right],
\label{eq:EinRay}
\end{equation}

\noindent 
where $l$ is the scattering length, $\lambda$ is the wavelength of light, $n$ is the index of refraction corresponding the wavelength of light, $T$ is temperature, $\rho$ is density, and $\kappa_{T}$ the isothermal compressibility. For this equation to be valid the index of refraction should be evaluated at the temperature, density and wavelength.~\cite{Land:1960}.  There are also material dependent correction factors than can be added to Equation~\ref{eq:EinRay} that do not apply in the case of nobles~\cite{More:1974A}. This expression for Rayleigh scattering length will be used in the extrapolations in section 3.

\subsection{Index of refraction}

The Sellmeier dispersion relation for liquids and solids at constant temperature and density \cite{Born:1999A, Schi:1990A}, is used to wavelength $\lambda$ to the index of refraction $n$.  This relation is

\begin{equation}
n^2 = a_0 + \sum_i\frac{a_i\lambda^2}{\lambda^2 - \lambda_i^2}.
\end{equation}

\noindent 
In this case $a_0$ is a Sellmeier coefficient that accounts for the effect of UV resonances not included in the sum and $a_i$ are the Sellmeier coefficients that correspond with the resonances, occurring at wavelength $\lambda_i$. The Sellmeier dispersion equation was derived from the Lorentz-Lorenz equation \cite{Loren:1880A, Land:1960} and the coefficients ($a_0$, $a_i$) are experimentally determined for a given medium. The scintillation wavelength of argon and xenon is between the UV and IR resonance peaks (shown in Table~\ref{tab:ArXeKrResonanceWL}), thus the following equation is sufficient for fitting the coefficients in the range of wavelengths around the scintillation wavelength around the scintillation wavelength,

\begin{equation}
\label{eq:sellfit}
n^2 = a_0 + \frac{a_{UV}\lambda^2}{\lambda^2 - \lambda_{UV}^2} + \frac{a_{IR}\lambda^2}{\lambda^2 - \lambda_{IR}^2},
\end{equation}

\noindent 
where $\lambda_{UV}$ corresponds to the closest or first UV resonance and $\lambda_{IR}$ corresponds to the closest or first IR resonance.

\begin{table}[ht]
\centering 
\begin{tabular}{| c | c c c |} 
\hline 
                 & Scintilation $\lambda$ & UV Resonance $\lambda$ & IR Resonance $\lambda$ \\ [0.5ex] 
Element	& ($nm$)                           & ($nm$)                                & ($nm$) \\
\hline 
Argon       & 128                                 & 106.6                                  & 908.3 \\
Xenon      & 178                                 & 146.9                                  & 827.0 \\
\hline
\end{tabular}
\caption{Scintillation and resonance wavelengths of argon and xenon. Argon UV resonance value obtained from~\cite{Lane:1967A}, xenon UV resonance obtained from~\cite{Bide:1981A} and argon and xenon IR resonances sourced from~\cite{Arai:1978A}. Xenon scintillation length obtained from \cite{Solo:2004A} and argon scintillation length obtained from \cite{Apri:2006A}.}
\label{tab:ArXeKrResonanceWL} 
\end{table}

\section{Calculation Expanding in Temperature and State}%%%%%%%%%%%%%%%%%%NEW!

In the calculations done by Seidel, to predict the index of refraction and scattering length in liquid nobles, the dispersion equation for gases was used to fit the gas data from \cite{Bide:1981A} for the Sellmeier coefficients. We update this calculation by fitting the Sellmeier coefficents (Equation~\ref{eq:sellfit}) to liquid and solid data from \cite{Sinn:1969A}, to find the respective scattering lengths in liquid and solid argon and xenon. The Sellmeier coefficients resulting from the fit can be found in Table~\ref{table:ArXeSellCo}

\begin{table}[H]

\centering 
\begin{tabular}{c c c c} 
\hline\hline 
T (K) & $a_0$ & $a_{UV}$ & $a_{IR}$\\ [0.5ex] 
\hline
\textit{Solid Argon} & & &\\
20 & 1.4$\pm$0.1 & 0.30$\pm$0.09 & 0.0011$\pm$0.007 \\
83.81 & 1.3$\pm$0.1 & 0.29$\pm$0.09 & 0.00087$\pm$0.007\\ 
\textit{Liquid Argon} & & &\\
83.81 & 1.24$\pm$0.09 & 0.27$\pm$0.09 & 0.00047$\pm$0.007\\
90 & 1.26$\pm$0.09 & 0.23$\pm$0.09 & 0.0023$\pm$0.007\\
\hline
\textit{Solid Xenon}& & &\\
80 & 1.6$\pm$0.3 & 0.6$\pm$0.2 & 0.001$\pm$0.03\\
162.35 & 1.4$\pm$0.2 & 0.6$\pm$0.2 & 0.0008$\pm$0.03\\ 
\textit{Liquid Xenon} & & &\\
162.35* & 1.5$\pm$0.02 & 0.38$\pm$0.01 & 0.009$\pm$0.01\\
178 & 1.4$\pm$0.2 & 0.4$\pm$0.2 & 0.002$\pm$0.02\\ 
\hline\hline 
\end{tabular}
\caption{Argon and Xenon Sellmeier coefficients calculated using data from \cite{Sinn:1969A}. $83.81~K$ is the argon triple point and $162.35~K$ is the xenon triple point. These coefficients are for cgs units. *This Fit includes the point from \cite{Solo:2004A}.} 
\label{table:ArXeSellCo} 
\end{table}

\subsection{Fit Verification}
Solovov \emph{et al.}~\cite{Solo:2004A} measured the index of refraction of the scintillation wavelength of liquid xenon at the triple point of temperature 162 $K$å. The liquid triple point data from~\cite{Sinn:1969A} was fit with equation~\ref{eq:sellfit} and used this fit to extrapolate to the scintillation wavelength. The extrapolation error was calculated using the covariance matrix, as described in~\cite{Tell:2001A}.  The predicted value is 1.69$\pm$0.04 predicts the measured value of 1.69$\pm$0.02. 

\subsection{Results}

\begin{figure} [H]
    \centering
   \includegraphics[width=4.0in]{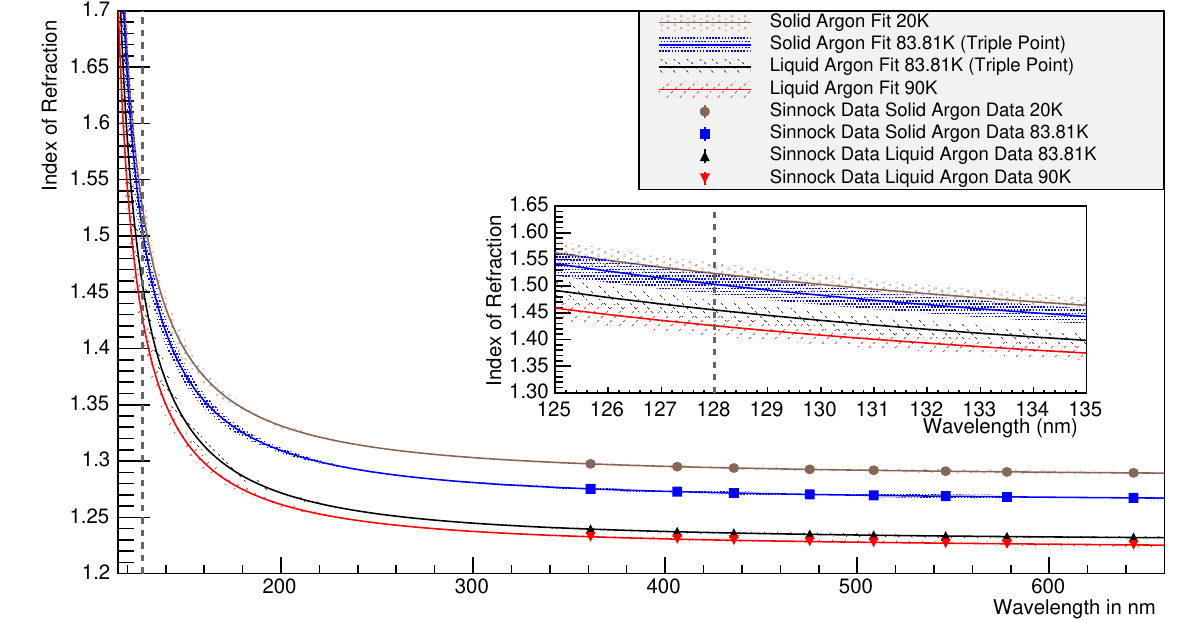}
   \caption{Calculated index of refraction vs. wavelength ($nm$) for solid and liquid argon.  The points show the data from reference \cite{Sinn:1969A}.}
   \label{fig:ArIorErr1}
\end{figure}

\begin{figure} [H]
    \centering
    \includegraphics[width=4.0in]{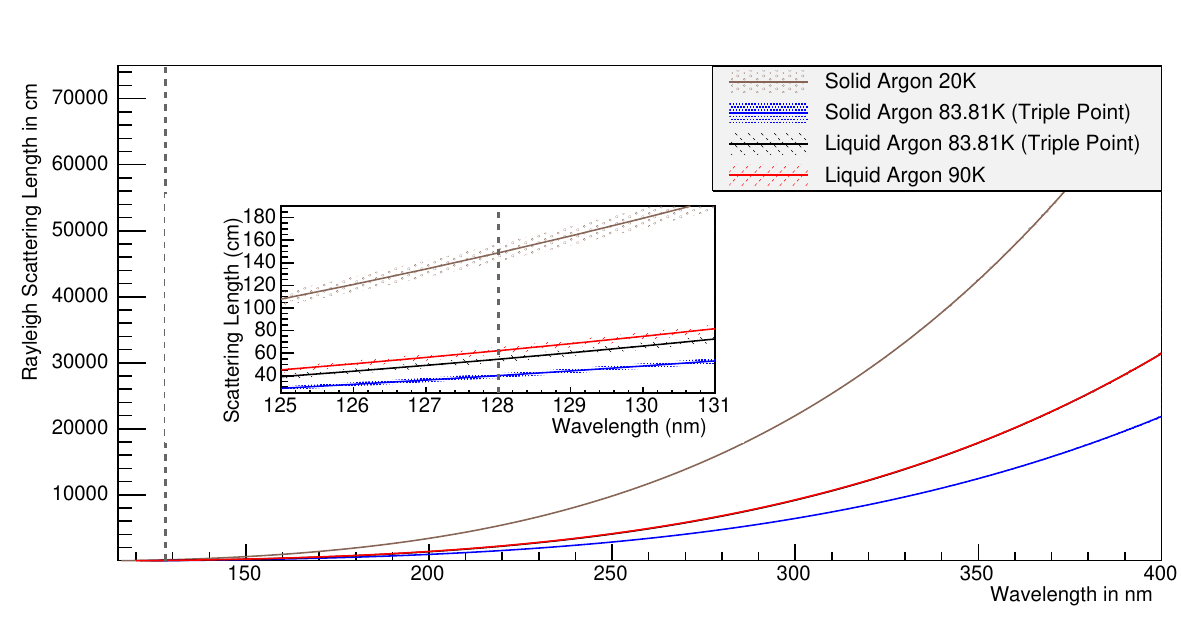}
    \caption{Calculated Rayleigh scattering length ($cm$) vs. wavelength ($nm$) for solid and liquid argon.}
    \label{fig:ArRSErr1}
\end{figure}

\begin{figure} [H]
    \centering
    \includegraphics[width=4.0in]{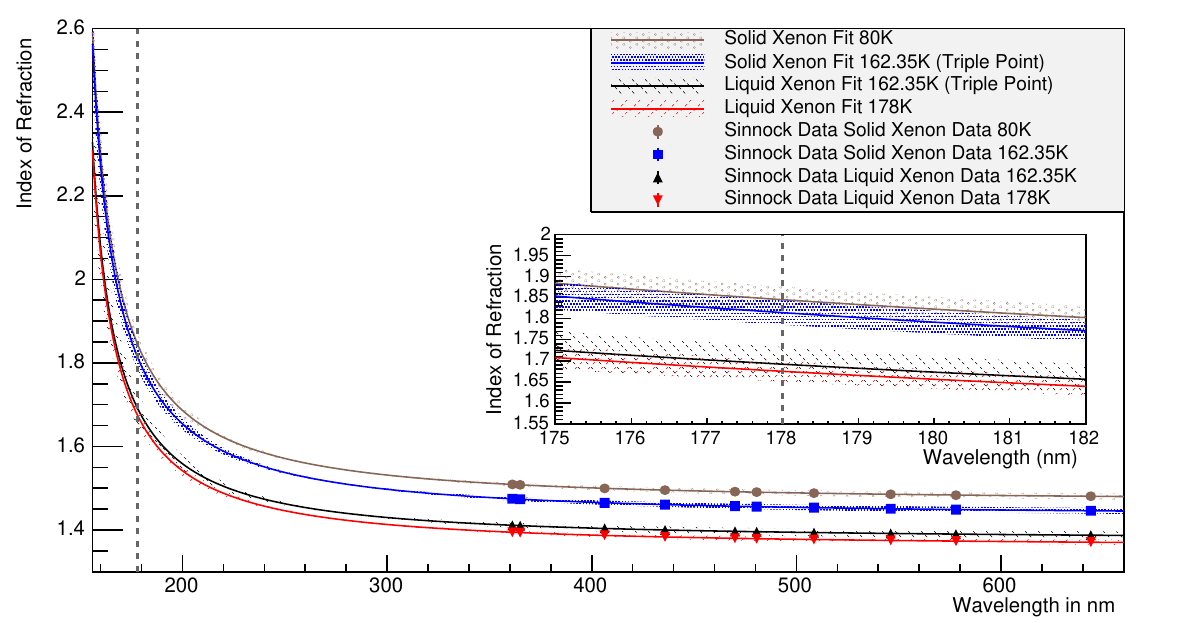}
    \caption{Calculated index of refraction vs. wavelength ($nm$) for solid and liquid xenon.  The points show the data from reference \cite{Solo:2004A} and \cite{Sinn:1969A}.}
    \label{fig:XeIorErr1}
\end{figure}

\begin{figure} [H]
    \centering
    \includegraphics[width=4.0in]{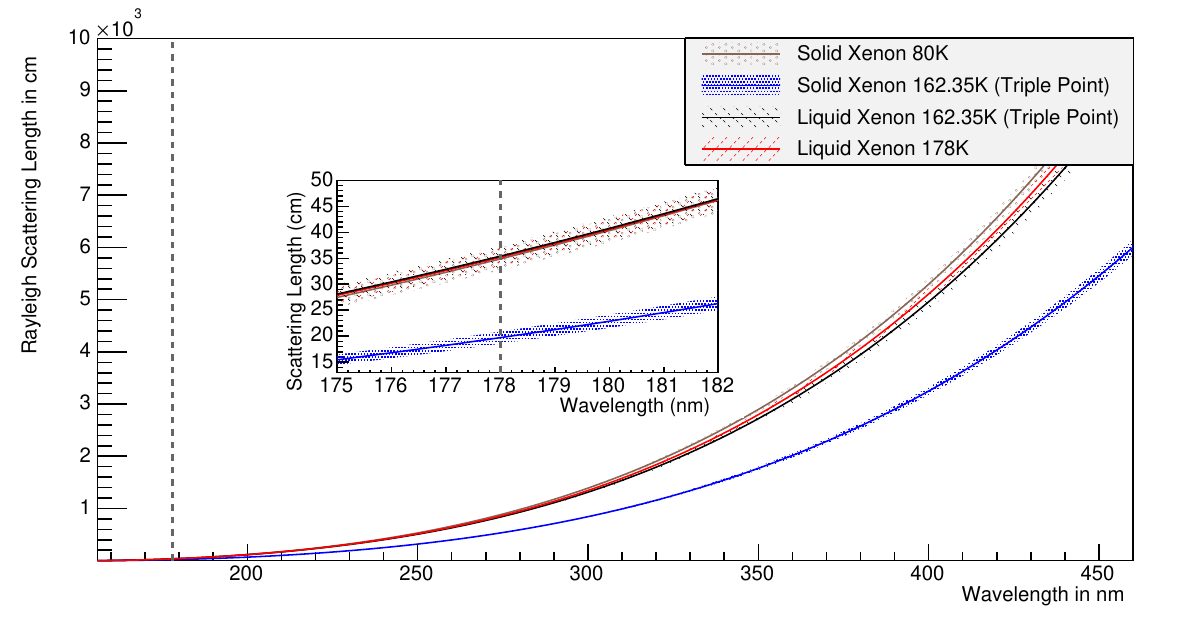}
    \caption{Calculated Rayleigh scattering length ($cm$) vs. wavelength ($nm$) for solid and liquid xenon.}
    \label{fig:XeRSErr1}
\end{figure}

The Sellemeier coefficients obtained by fitting the data from Sinnock and Smith~\cite{Sinn:1969A} were then used to extrapolate the index of refraction and Rayleigh scattering length at the scintillation wavelengths. The new calculated values of the index of refraction and Rayleigh scattering length vs. wavelength are shown in Figures \ref{fig:ArIorErr1} and \ref{fig:ArRSErr1} for liquid argon, and in Figures \ref{fig:XeIorErr1} and \ref{fig:XeRSErr1} for xenon in both solid and liquid phases. The results of this new calculation are summarized in Table \ref{tab:FinalResults}.

\begin{table}[ht]
\centering 
\begin{tabular}{c | c c c | c c c} 
\hline\hline 
&  & n &  &  & l ($cm$) &  \\ [0.5ex] 
                                   & {\textbf{This}}    & {Previous}     & {Previous}             & {\textbf{This}} & {Previous}    & {Previous}\\
Element                      & {\textbf{Calculation}}    & {Calculation} & {Measurement}     & {\textbf{Calculation}} & {Calculation} & {Measurement}\\
\hline 
Liquid Argon               & $1.45\pm0.07$        & 1.37              & N/A                        & 55$\pm$5             & 90                 &  $66\pm3$ \\
Liquid Xenon              & $1.69\pm0.04$        & 1.68              & 1.69$\pm$0.02      & 35$\pm$2             &  30                &  $36\pm2$ \\
Solid Argon                 & $1.50\pm0.07$        & N/A               & N/A                        & 40$\pm$4             &  N/A              &  N/A           \\
Solid Xenon                & $1.81\pm0.03$        & N/A               & N/A                        & 20$\pm$1             &  N/A              &  N/A           \\
\hline\hline 
\end{tabular}
\caption{This a summary of  the results of the extrapolations made by fitting the Sinnock data at the triple point with the Sellmeier equation. Both the liquid and solid triple point values are included. These values are compared with previous calculations and measurements. The previous index of refraction and Rayleigh scattering length calculation come from~\cite{Seid:2002A}; error bars were not included in the original work. The previous argon scattering length measurements come from~\cite{Ishi:1997A} and the previous xenon index of refraction measurement and attenuation length is from~\cite{Solo:2004A}.} 
\label{tab:FinalResults} 
\end{table}

\section{Conclusion}

This calculation has produced the first prediction for the Rayleigh scattering length in liquid argon and xenon for which the Sellmeier dispersion constants are fit from measurements at the same temperature and state.  This is the first calculation of the scattering length in solid argon and xenon. This analysis used the data from Sinnock \emph{et al.}~\cite{Sinn:1969A} to extrapolate the wavelength dependent index of refraction through argon at constant temperature and state. Using this index of refraction, the Rayleigh scattering length in liquid argon at 90 $K$ is calculated to be 60$\pm$6 $cm$ and 55$\pm$5 $cm$. The extrapolation of the index of refraction at 90 K is within error of the value measured by Ishida \emph{et al.}~\cite{Ishi:1997A} unlike the extrapolated value by Seidel \emph{et al.}~\cite{Seid:2002A}. The accuracy of the extrapolation method was tested against a measured point in xenon and we were able to predict the value within experimental error. 

ArDM used the results of this calculation to update the optical simulation. Using this, an in situ measurement of the attenuation length was made with result of 52$\pm$7cm \cite{Calv:2016A}. This is within error the predicted value.

The data taken by Sinnock \emph{et al.}~\cite{Sinn:1969A} also gave us the opportunity to produce values for the index of the refraction and Rayleigh scattering lengths of solid argon and xenon at different temperatures at the scintillation wavelengths. These values may be useful in future detectors or experiments that take advantage of the scintillation properties of these elements in a solid state.  All of the results for the index of refraction and Rayleigh scattering length are collected in table~\ref{tab:FinalResults}.

\section{Acknowledgements}
We acknowledge support from the European Research Council Project ERC StG 279980, the UK Science and Technology Facilities Council (STFC) grant ST/K002570/1, the Leverhulme Trust grant number ECF-20130496, and STFC and SEPNet PhD studentship support.

\bibliography{RSCalc}
\end{document}